# Observations of gravity wave forcing of the mesopause region during the January 2013 major Sudden Stratospheric Warming

R. J. de Wit[1], R. E. Hibbins[1,2], P. J. Espy[1,2], Y. J. Orsolini[2,3], V. Limpasuvan[4], and D. E. Kinnison[5]

[1]Department of Physics, Norwegian University of Science and Technology, Trondheim, Norway, [2]Birkeland Centre for Space Science, Bergen, Norway, [3]Norwegian Institute for Air Research, Kjeller, Norway, [4]School of Coastal and Marine Systems Science, Coastal Carolina University, Conway, South Carolina, USA, [5]National Center for Atmospheric Research, Boulder, Colorado, USA

**Abstract** Studies of vertical and interhemispheric coupling during Sudden Stratospheric Warmings (SSWs) suggest that gravity wave (GW) momentum flux divergence plays a key role in forcing the middle atmosphere, although observational validation of GW forcing is limited. We present a whole atmosphere view of zonal winds from the surface to 100 km during the January 2013 major SSW, together with observed GW momentum fluxes in the mesopause region derived from uninterrupted high-resolution meteor radar observations from an All-Sky Interferometric Meteor Radar system located at Trondheim, Norway (63.4°N, 10.5°E). Observations show GW momentum flux divergence 6 days prior to the SSW onset, producing an eastward forcing with peak values of $\sim$+145 ± 60 m s$^{-1}$ d$^{-1}$. As the SSW evolves, GW forcing turns westward, reaching a minimum of $\sim$−240 ± 70 m s$^{-1}$ d$^{-1}$ $\sim$+18 days after the SSW onset. These results are discussed in light of previous studies and simulations using the Whole Atmosphere Community Climate Model with Specified Dynamics.

## 1. Introduction

In the Northern Hemisphere winter stratosphere dramatic dynamic events known as Sudden Stratospheric Warmings (SSWs) occur roughly every other year [*Gerber et al.*, 2012], during which the polar vortex reverses and stratospheric polar temperatures rapidly increase. SSWs are understood to develop as a result of the interaction of bursts of upward propagating planetary waves (PWs) with the stratospheric zonal wind [*Matsuno*, 1971]. Despite the name, SSWs exert dynamical effects beyond the stratosphere, extending all the way from the surface up to the mesosphere and ionosphere [e.g., *Baldwin and Dunkerton*, 2001; *Hoffmann et al.*, 2007; *Goncharenko et al.*, 2010].

The high-latitude mesosphere/lower thermosphere (MLT) response to SSWs is characterized by a weakening or reversal of the climatological eastward winds, with the reversal in the MLT generally preceding the SSW onset in the lower stratosphere [e.g., *Hoffmann et al.*, 2002, 2007; *Mukhtarov et al.*, 2007]. Enhanced PW activity in the MLT [e.g., *Jacobi et al.*, 2003; *Hoffmann et al.*, 2007], as well as mesospheric cooling [*Hoffmann et al.*, 2007] associated with SSWs, has also been reported.

In addition to the PW drivers, atmospheric gravity waves (GWs), the main driver of the general circulation of the quiescent mesosphere, appear to play an important role in coupling the different atmospheric layers during SSWs [e.g., *Holton*, 1983; *Zülicke and Becker*, 2013]. Stratospheric zonal winds are disturbed during SSW conditions, changing filtering conditions for upward propagating GWs, resulting in altered GW fluxes and forcing in the mesosphere. Mesospheric cooling associated with SSWs, for example, has been attributed to a weakening of the westward GW forcing due to the disturbed stratospheric wind conditions. This results in reduced downwelling and a relaxation of mesospheric temperatures toward radiative equilibrium [*Holton*, 1983].

Although stratospheric behavior during SSWs is generally well characterized and understood both from an observational and modeling perspective [*Hoffmann et al.*, 2002; *Mukhtarov et al.*, 2007], characterization and understanding of the MLT response is still incomplete [*Hoffmann et al.*, 2002; *Mukhtarov et al.*, 2007; *Smith*, 2012]. GWs are known to play an important role in MLT dynamics, but their evolution during SSWs is not well understood, and a better knowledge of GW propagation during SSWs is necessary to correctly model vertical and interhemispheric coupling of the atmosphere in the MLT [*Yamashita et al.*, 2010; *Becker and Fritts*, 2006]. While GW variance has been used to study the variability of GWs during SSWs [e.g., *Hoffman et al.*, 2007; *Yamashita et al.*, 2013], observations of GW momentum flux and GW forcing during SSWs are not available. Such observations are required to provide GW directional information (available from GW momentum flux measurements), as GW variance is a scalar quantity lacking information about GW propagation direction. In addition, the directional forcing can be directly compared to modeling results.

In this study, a whole atmosphere view of zonal winds in the troposphere, stratosphere, and mesosphere during the 2012–2013 major SSW, with an onset on 7 January 2013, is presented. In addition, we report for the first time observations of both momentum flux and forcing due to GWs in the MLT during the evolution of the SSW derived from meteor radar observations. These observations are discussed in light of previous observations and the Whole Atmosphere Community Climate Model with Specified Dynamics (WACCM-SD) simulations for the 2012–2013 winter.

## 2. Data and Analysis

Vertical profiles of MLT winds and GW momentum fluxes have been derived from near-continuous meteor radar measurements over Trondheim, Norway (63.4°N, 10.5°E), between December 2012 and February 2013. The time series is largely uninterrupted, with only three data gaps, each lasting 4 h or less, present on 31 January, 12 February, and 18 February. The Trondheim meteor radar has been operational since September 2012 and is a new generation All-Sky Interferometric Meteor Radar (SKiYMET) [*Hocking et al.*, 2001]. The system consists of eight transmitting antennas, phased to transmit most of the power at zenith angles between 15° and 50° and as such is optimized to measure high-frequency GW momentum fluxes. The radar is similar in design to the SAAMER and DrAAMER systems [*Fritts et al.*, 2010, 2012], with the Trondheim radar operating at 34.21 MHz with a 30 kW peak power. We observe maximum meteor count rates around 90 km and average daily count rates of around 6500 unambiguous meteors between 70–100 km and 15–50° zenith angle during the period under consideration. A detailed description of the Trondheim system and first year of observations will be detailed in a subsequent publication (in preparation).

Between 70 and 100 km, 3-hourly averaged horizontal winds, stepped in hourly intervals, have been determined from measured radial velocities [*Hocking et al.*, 2001], using altitude bins of 5 km (70–80 km), 4 km (80–84 km), 2 km (84–96 km), and again 4 km (96–100 km) to correct for the decrease of meteor counts away from 90 km. Throughout all analyses described in this study only unambiguous meteors detected between 15° and 50° zenith angle have been used. These hourly wind values were then least squares fitted over an interval of 4 days (time stepped by 1 day) using an offset (that represented the 4 day moving average horizontal wind), as well as oscillations with periods of 48 (2 day wave), 24, 12, and 8 h (tides).

Below 68 km the meteor radar wind data have been complemented by Modern Era Retrospective Analysis for Research and Applications (MERRA) reanalyses [*Rienecker et al.*, 2011] at 63.4°N and 10.5°E in order to study the zonal wind development over the full vertical range of the neutral atmosphere, from the surface to 100 km. Zonal winds are available 4 times daily, of which 4 day moving averages (time stepped by 1 day) have been created. The zonal mean of the 4 day moving average zonal wind reversed direction at 10 hPa, 60°N on 7 January 2013. This date will be referred to as the "onset" of the SSW throughout this study.

The high-frequency zonal GW momentum fluxes, $\overline{u'w'}$, where the primed quantities denote a deviation from the background state (assumed to be caused by high-frequency GWs), have been determined using a 10 day moving average for four 5 km bins between 80 and 100 km. *Andrioli et al.* [2013] pointed out that spurious signals can arise due to the incomplete removal of background winds, especially the large-amplitude high-frequency tides commonly seen at these latitudes. Therefore, the determination of the background wind as well as the mean meteor height and time was performed using fine temporal (1 h) and vertical (2 km (82–96 km) and 4 km (78–82 km, 96–100 km)) resolution. This resolution choice was found to minimize the variance in the residual winds. In order to correct for vertical and temporal shears in the background wind within a given time and height interval (bounded by the mean meteor height and time), the background wind is linearly interpolated between adjacent intervals to the time and height of each individual meteor echo within the given interval. The component of this value along the meteor line of sight is subsequently subtracted off the individual meteor's observed radial velocity to derive the residual velocity perturbation due to GWs. Hourly momentum fluxes are then calculated from these residual perturbation velocities when at least 30 meteors are available, using the matrix-inversion method outlined in *Hocking* [2005]. When calculating the 10 day moving average momentum fluxes, results where $|\overline{u'w'}| > 300$ m$^2$ s$^{-2}$, or where the matrix to be inverted is near singular, are discarded. Although these results are mathematically correct, they are considered nonphysical and hence are not taken into account in any further analysis (W.K. Hocking, personal communication, 2013).

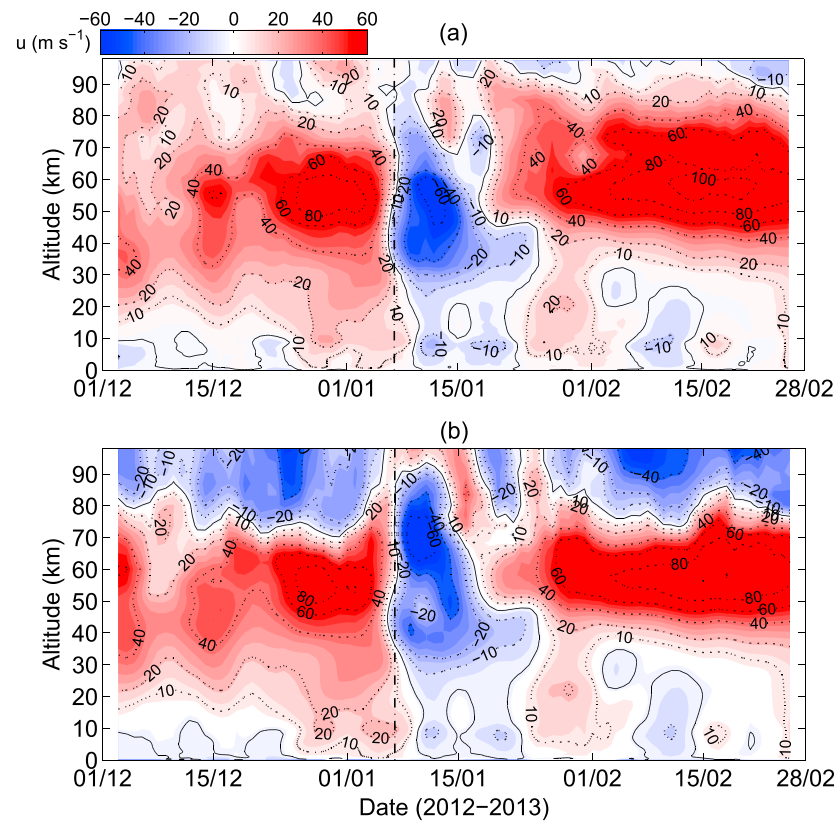

**Figure 1.** Four day moving average zonal wind (in m s$^{-1}$) over Trondheim, Norway, (a) derived from meteor radar observations (70–100 km) complemented with MERRA reanalysis results (below 68 km) and (b) as simulated in WACCM-SD for December 2012 to February 2013. Highlighted with dotted lines are ±10, ±20, ±40, ±60, ±80, and ±100 m s$^{-1}$. Zero contour and onset of SSW are indicated by a black line and black dash-dotted line, respectively. White spaces around the start and end dates are caused by the creation of a 4 day moving average using data ranging from 1 December 2012 to 28 February 2013.

The quantity $\overline{u'w'}$ gives information about GW propagation direction and strength. However, to study the high-frequency GW forcing (GWF) the vertical divergence of $\overline{u'w'}$, corrected for the decrease in density with height must be used. Ten day moving averages of the density corrected momentum flux, $\rho\overline{u'w'}$, for the altitude intervals 80–90 km and 90–100 km, have been derived using CIRA-86 [*Fleming et al.*, 1990] monthly mean densities. The forcing caused by high-frequency GWs can then be determined from $\rho\overline{u'w'}$ using $GWF = -\frac{1}{\rho}\frac{\partial(\rho\overline{u'w'})}{\partial z}$ [e.g., *Fritts and Vincent*, 1987]. Using $\overline{u'w'}$ calculated for the 80–90 km and 90–100 km height intervals, GWF at ~90 km can be estimated.

To further investigate these processes, the Trondheim observations for the 2012–2013 SSW are compared to a WACCM Specified Dynamics (WACCM-SD) simulation, in which the dynamics and temperature are nudged in the lower part of the model domain by MERRA reanalysis data [*Rienecker et al.*, 2011]. Specifically, MERRA temperature, zonal and meridional winds, and surface pressure are used to drive the physical parameterizations that control boundary layer exchanges, advective and convective transport, and the hydrological cycle within WACCM. The WACCM meteorological fields are constrained by the MERRA meteorological fields using the approach described in *Kunz et al.* [2011]. This constraint is applied at every model time step (i.e., every 30 min). Consistent with a 50 h relaxation time constant, 1% of the MERRA fields were used at every time step. For the meteorological fields, this scheme is applied from the surface to approximately 50 km (0.79 hPa); above 60 km (0.19 hPa) the WACCM meteorological fields are unconstrained and fully interactive, with a linear transition in between.

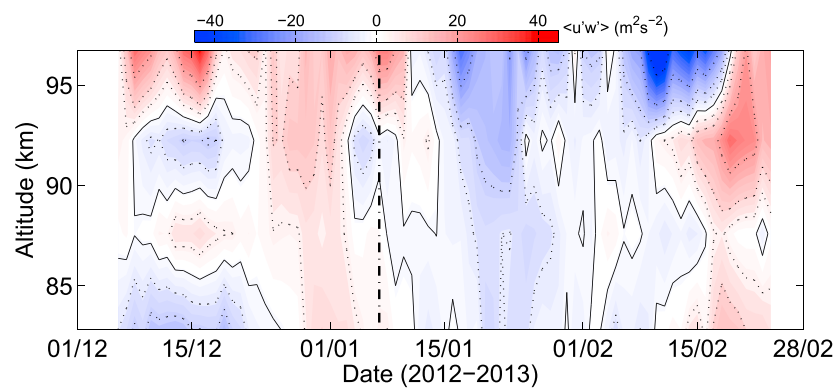

**Figure 2.** Ten day moving average vertical flux of zonal momentum due to high-frequency GWs (in m$^2$ s$^{-2}$). Levels at ±5, ±10, ±20, and ±40 m$^2$ s$^{-2}$ are highlighted with dotted lines. Zero contour and onset of SSW are indicated by a black line and a black dash-dotted line, respectively. White spaces around the start and end dates are caused by the creation of a 10 day moving average using data ranging from 1 December 2012 to 28 February 2013.

### 3. Results

The 4 day moving average zonal wind observed over Trondheim, Norway,

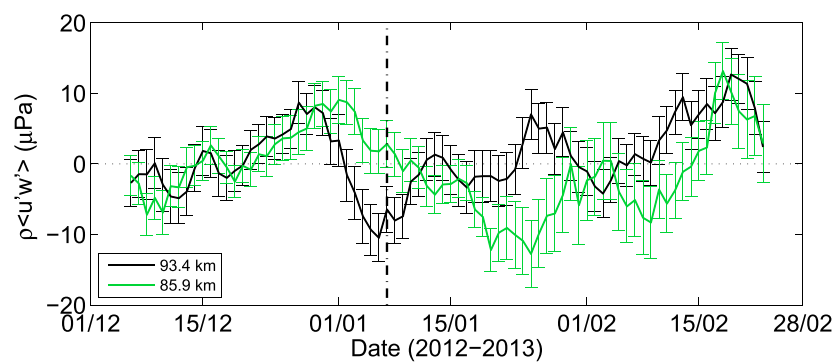

**Figure 3.** Ten day moving average density weighted vertical flux of zonal momentum due to high-frequency GWs (in µPa) for 80–90 km (green) and 90–100 km (black). Onset of the SSW is indicated by a black dash-dotted line.

from the surface to 100 km is shown in Figure 1a for December 2012 to February 2013. Throughout December, typical winter behavior can be observed: the eastward winter vortex maximizes in the upper stratosphere/lower mesosphere, reaching values of just under 100 m s$^{-1}$ during the second half of December. Above this region, the zonal wind decreases in strength although remaining predominantly eastward.

Toward the end of December the eastward winds start to weaken, and a rapid wind reversal takes place at all altitudes from the lower troposphere to 100 km around 7 January, in this study defined as the onset of the major SSW (see section 2). Between ∼35–55 km and ∼75–85 km, the wind reversal precedes the onset, whereas above 85 km the reversal lags by some days. The disturbed conditions last for several days, after which the winds slowly return to eastward again, starting from the upper levels. Post-SSW winds between 70 and 85 km are more strongly eastward in comparison to both pre-SSW (Figure 1a) and climatological conditions over Scandinavia [*Sandford et al.*, 2010].

For comparison with the observed winds in Figure 1a, the 4 day moving average zonal wind over Trondheim derived from WACCM-SD is shown in Figure 1b. Generally, the model winds show behavior similar to the observed winds, although the weakening of the winds above the upper stratosphere/lower mesosphere in December and February is stronger than observed. This results in larger vertical wind shear, and a turning of the model winds to westward above about 70–80 km, as compared to weak eastward winds observed at these altitudes.

The observed 10 day moving average vertical flux of zonal momentum due to high-frequency GWs, $\overline{u'w'}$, is presented in Figure 2. The most notable feature is the prolonged period of westward momentum flux starting during the second half of January, reaching values of up to −50 m$^2$ s$^{-2}$ and coinciding with the enhanced eastward winds between 70 and 85 km. In contrast, just prior to the SSW onset, when winds around 60 km start to weaken, the zonal momentum flux is observed to be eastward throughout the full altitude range under consideration, with the exception of a small region of westward momentum flux between ∼87 and 93 km just before the onset.

To derive the zonal GWF around 90 km, 10 day moving averages of the density weighted momentum flux, $\rho\overline{u'w'}$, for the altitude intervals 80–90 km (Figure 3, in green) and 90–100 km (in black), have been derived.

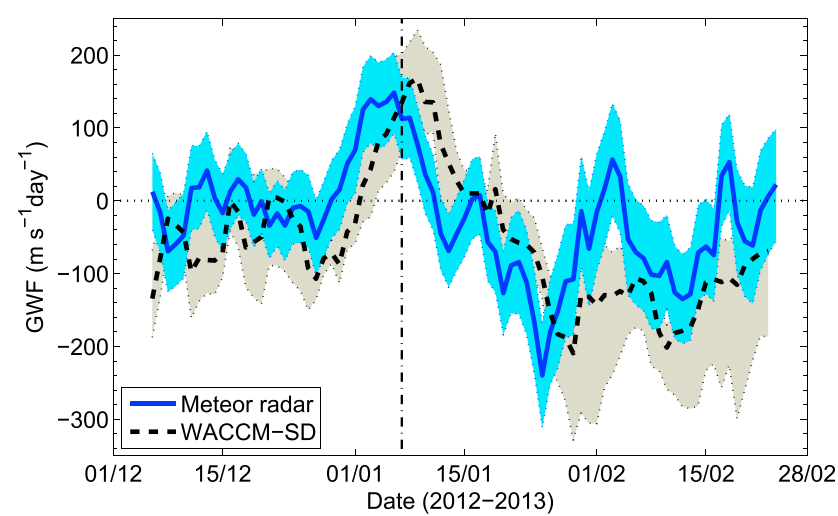

**Figure 4.** Ten day moving average GWF (in m s$^{-1}$ d$^{-1}$) at 90 km over Trondheim, Norway, derived from meteor radar observations (blue, errors are shaded) and as simulated in WACCM-SD (black dashed line, area between maximum and minimum GWF in layer is shaded). Onset of the SSW is indicated by a black dash-dotted line.

The derived values are in the range of ±20 µPa and are of the same order of magnitude as those observed in the high-latitude Northern Hemisphere MLT region in January 2006 by *Ern et al.* [2011], who used Sounding of the Atmosphere Using Broadband Emission Radiometry (SABER) data to derive directionless density weighted momentum fluxes. Comparing $\rho\overline{u'w'}$ for the two height intervals, a significant divergence is seen around the SSW and again around 2–3 weeks and about 5 weeks after the SSW onset.

From the density weighted momentum fluxes the forcing due to high-frequency GWs is derived, as shown in Figure 4 (in blue). Until about 1 week before the SSW onset the GWF is not statistically significantly different from zero, as expected from the overlaying $\overline{\rho u'w'}$ profiles during this time period. Coincident with the weakening of the eastward winds, the GWF turns eastward reaching peak values of ∼+145 ± 60 m s$^{-1}$ d$^{-1}$ days before the SSW onset. After the SSW, from mid-January onward when eastward winds below 85 km are enhanced, the GWF at 90 km is predominantly westward reaching peak values of ∼−240 ± 70 m s$^{-1}$ d$^{-1}$ around 18 days after the SSW onset and returning to pre-SSW values toward the end of February.

## 4. Discussion

Many observational studies have reported a mesospheric wind reversal prior to the SSW onset [e.g., *Hoffmann et al.*, 2002, 2007; *Matthias et al.*, 2012], whereas post-SSW an enhancement of mesospheric eastward winds has been noted [*Hoffmann et al.*, 2007; *Orsolini et al.*, 2010]. In the present study, the zonal wind reversal between about 75 and 85 km precedes the SSW onset by 1 to 2 days, and post-SSW zonal wind behavior is also characterized by enhanced eastward winds below 85 km. In contrast, the wind reversal above about 85 km over Trondheim lags the SSW onset. It should be noted that the effect of the SSW may depend on the position of the vortex relative to the observational sites [*Jacobi et al.*, 2003]. Thus, comparisons of reversal times between stations will be highly dependent upon the particular dynamics occurring above each station. Furthermore, it should be noted that due to the use of a 4 day moving average time shifts of 1 to 2 days should be interpreted with care.

Coincident with the enhanced eastward winds after the SSW, *Hoffmann et al.* [2007] observed an increase of turbulent energy dissipation rates and GW activity. Our results indicate increased GW activity during the period of enhanced post-SSW eastward winds as well, with maximum westward GW momentum fluxes and forcing occurring at this time. The westward $\overline{u'w'}$ can be readily explained by selective filtering [*Lindzen*, 1981] of upward propagating eastward GWs by the layer of increased eastward winds below 85 km (cf. Figure 1), resulting in the removal of eastward propagating waves and subsequent increased net westward $\overline{u'w'}$ in the MLT. Moreover, GWF is observed to turn westward in the period during which this increased westward momentum flux is observed. Similarly, when eastward stratospheric winds start to weaken around the end of December, the GWF turns eastward, peaking just before the SSW onset.

The occurrence of peak eastward GWF before the SSW onset and hence before the maximum westward stratospheric winds, cannot be simply explained by selective filtering of a *uniform* spectrum of vertically upward propagating GWs by the underlying zonal winds. However, it should be noted that the GW spectrum may be skewed to include more eastward than westward propagating GWs breaking in the mesopause region prior to the SSW, causing the eastward GWF to peak at this time. Furthermore, ray-tracing simulations performed by *Yamashita et al.* [2013] have shown that during certain SSW events GWs can propagate meridionally from the midlatitudes to higher latitudes. However, a full discussion of the spectrum and source region of the GWs causing the observed GWF is beyond the scope of this study.

The corresponding GWF from WACCM-SD is displayed in Figure 4 (indicated by the dashed black line). Here the 10 day moving average GWF time series is derived from GWF averaged between ∼80 and 100 km and over an areal region centered over Trondheim (from 61 to 65°N, 7.5 to 12.5°E). The GWF is derived from GWs generated by tropospheric fronts at 600 hPa, where their source spectra are specified [*Richter et al.*, 2010]. As noted by *Limpasuvan et al.* [2012], GWF produced by orographic and convection sources is not important in the polar region.

As in nearly all climate models, the GW parameterization in WACCM is based on the assumption that GWs, once launched, only propagate upward vertically before depositing their momentum. The column-by-column implementation of GWs dictates an areal average around Trondheim to reduce noise in the presented data. Additionally, vertical averaging is performed to mimic the observational derivation of GWF, in which the momentum flux divergence between two layers (80–90 km and 90–100 km) is computed.

The observational and model GWF results show qualitatively very similar behavior, increasing confidence in both techniques. Prior to the SSW, the observed GWF is around 0 m s$^{-1}$ d$^{-1}$, while the modeled GWF is weakly negative. Like the observed GWF, just before the onset of the SSW the WACCM GWF becomes eastward, lasting until several days after the onset. While the peak eastward forcing amplitudes are comparable (∼150 m s$^{-1}$ d$^{-1}$), there is a 3 day time difference between the GWF peaks. This slight time difference in the GWF peak may be attributable to observational uncertainty, meridional propagation of GWs from lower latitude (not parameterized in WACCM), forcing due to in situ generation of GWs by secondary wave breaking [e.g., *Satomura and Sato*, 1999] or by spontaneous adjustments of the vortex from below [e.g., *Sato and Yoshiki*, 2008; *Limpasuvan et al.*, 2011], as well as the zonal mean wind difference between WACCM and observations at the altitudes not constrained by MERRA (Figure 1). Post-SSW the GWF is westward in both model and observations, reaching peak values of $\sim -200$ m s$^{-1}$ d$^{-1}$.

Despite the difference in the zonal winds above 80 km, the similarity between the GWF derived from the meteor radar observations and WACCM-SD simulations is striking. This suggests that the asymmetry in the breaking gravity waves at 90 km is primarily driven by the critical level wave filtering below 90 km rather than the direction and strength of the background wind in which the waves break. Investigations of the spectrum of the gravity waves breaking at 90 km and their longitudinal variability during the evolution of an SSW are currently underway and will form the basis of a future paper.

Qualitatively, the GWF presented for the January 2013 SSW is comparable to those shown in previous modeling studies. *Chandran et al.* [2011], in a case study using free-running WACCM simulations, showed eastward GWF of up to 70 m s$^{-1}$ d$^{-1}$ (polar cap zonal mean) to occur at around 90 km in association with major SSWs, whereas during dynamically quiet winter times GWF was seen to be weakly westward. Similar results were obtained by *Limpasuvan et al.* [2012], who found a turning of GWF from westward to eastward before the onset of the SSW, and a decrease to climatological values after the SSW. *Miller et al.* [2013] showed an eastward GWF anomaly around the SSW date, and alternating eastward and westward anomalies from 6 to 15 days after the wind reversal at $\sim$90 km at 60°N using a composite of internally generated SSWs in a 20 year Hamburg Model of the Neutral and Ionized Atmosphere run. *Zülicke and Becker* [2013], using a composite of five SSWs internally generated in the Kühlungsborn Mechanistic Circulation Model, derived eastward GWF around the SSW at around 0.0015 hPa between 60 and 70°N (their Figure 6c). However, the peak as observed over Trondheim is approximately an order of magnitude higher than the forcing found in these last two modeling studies. This could be related to the spatial averaging used in displaying the model results.

The circulation in the MLT region is understood to be largely driven by GWF. Post-SSW the observed GWF is mainly westward as expected during normal winter conditions [*Körnich and Becker*, 2010], exerting a drag on the eastward winds and resulting in the observed decrease of zonal wind with height. On the other hand, the MLT westward winds during the SSW cannot be simply explained by the eastward GWF around 90 km during this time. In a case study with the free-running WACCM model *Limpasuvan et al.* [2012] showed that during the SSW much of the GWF is negated by strong PW forcing above 80 km. Using a composite of 54 major SSWs, *Chandran et al.* [2013] confirmed this behavior. Observations have indeed shown increased MLT PW activity associated with SSWs [*Jacobi et al.*, 2003]. It should be noted that during the January 2013 SSW waves with periods of around 12 days and amplitudes up to 30 m s$^{-1}$ were present in the meridional winds in the MLT above Trondheim (not shown). Although the forcing due to these long-period PWs may offset the high-frequency GWF, a full discussion of these effects is beyond the scope of the current study.

## 5. Conclusions

For the first time observations of high-frequency GW momentum fluxes and GWF during a major SSW have been reported. The GW momentum flux and forcing data, presented for December 2012 to February 2013, are obtained with a new generation SKiYMET meteor radar at Trondheim, Norway (63.4°N, 10.5°N). Together with the high-frequency GW momentum fluxes and GWF, a whole atmospheric view of the tropospheric, stratospheric, and mesospheric zonal winds over Trondheim has been presented by combining meteor radar data between 70 and 100 km with MERRA reanalyses. A rapid reversal of eastward to westward winds associated with the SSW is observed from the surface to 100 km. Post-SSW eastward winds between 70 and 85 km are enhanced, resulting in selective filtering of eastward GWs, causing the observed westward GW momentum fluxes of up to $-50$ m$^2$ s$^{-2}$. GWF starts to increase from zero around 1 week before the SSW onset, when stratospheric eastward winds start to weaken, reaching peak values of $\sim +145 \pm 60$ m s$^{-1}$ d$^{-1}$ several days before the SSW onset. Post-SSW, when the eastward winds below 85 km are enhanced, the GWF is generally westward with peak values of $\sim -240 \pm 70$ m s$^{-1}$ d$^{-1}$, returning to pre-SSW values toward the end of February. Furthermore, comparison of observed GWF with WACCM-SD results shows good qualitative agreement, increasing confidence in both methods.

**Acknowledgments**

MERRA results are available through http://disc.sci.gsfc.nasa.gov/daac-bin/DataHoldings.pl. Information on obtaining WACCM source code can be found on https://www2.cesm.ucar.edu/working-groups/wawg/code. Trondheim meteor radar data are available from the authors upon request. This study was partly supported by the Research Council of Norway/CoE under contract 223252/F50. V.L. was supported by the National Science Foundation (NSF) under grants AGS-1116123 and AGS-MRI-0958616 and the South Carolina NASA Space Grant.

The Editor thanks Vânia Fátima Andrioli and an anonymous reviewer for their assistance in evaluating this paper.